\documentclass[%
preprint,
 amsmath,amssymb,
 aps,
prb,
]{revtex4-2}

\usepackage{graphicx}
\usepackage{dcolumn}
\usepackage{bm}
\usepackage [english]{babel}
\usepackage [autostyle, english = american]{csquotes}
\usepackage{subfiles}
\usepackage{textgreek}

\begin{document}


\title{Crystal-like thermal transport in amorphous carbon}

\author{Jaeyun Moon}
\email{To whom correspondence should be addressed; E-mail: jaeyun.moon@cornell.edu}
 \affiliation{Sibley School of Mechanical and Aerospace Engineering, Cornell University}
\author{Zhiting Tian}
\affiliation{Sibley School of Mechanical and Aerospace Engineering, Cornell University}

\date{\today}
\clearpage

\begin{abstract}
Thermal transport properties of amorphous carbon has attracted increasing attention due to its extreme thermal properties: It has been reported to have among the highest thermal conductivity for bulk amorphous solids up to $\sim$ 37 Wm\textsuperscript{-1}K\textsuperscript{-1}, comparable to crystalline sapphire ($\alpha$-Al\textsubscript{2}O\textsubscript{3}). Further, large density dependence in thermal conductivity demonstrates a potential for largely tunable thermal conductivity. However, mechanism behind the high thermal conductivity and its large density dependence remains elusive due to many variables at play. In this work, we perform large-scale ($\sim$ 10\textsuperscript{5} atoms) molecular dynamics simulations utilizing a machine learning potential based on neural networks. Through spectral decomposition of thermal conductivity which enables a quantum correction to classical heat capacity, we find that propagating vibrational excitations govern thermal transport in amorphous carbon ($\sim$ 100 \% of thermal conductivity) in sharp contrast to the conventional wisdom that diffusive vibrational excitations dominate thermal transport in amorphous solids. Instead, this remarkable behavior resembles thermal transport in simple crystals. Moreover, our temperature dependent spectral diffusivity and velocity current correlation analyses reveal that the density dependent thermal conductivity originates from anharmonicity sensitive propagating excitations. Our work suggests a novel insight and design principle into developing mechanically hard, thermally conductive amorphous solids. 

\end{abstract}


\maketitle
\clearpage

\section{Introduction}

Vibrational properties of amorphous solids are of fundamental interest due to their anomalies compared to those of crystalline solids, including a Boson peak in the vibrational density of states \cite{shintani_universal_2008, chumakov_role_2014} and an excess heat capacity at cryogenic temperatures \cite{zeller_thermal_1971}. Combined with small average mean free paths on the order of interatomic distances \cite{kittel_interpretation_1949}, these anomalies suggest different characteristics of vibrational heat carriers from those of phonons in simple crystals. Prior seminal works have proposed to categorize these heat carriers in amorphous solids as propagating, diffusive, and localized vibrations depending on their transport mechanisms \cite{allen_thermal_1989, allen_diffusons_1999, allen_character_2003}. Propagating vibrational excitations are similar to phonons in that they have large mean free paths, thus transporting heat efficiently. Diffusive vibrations have ill-defined mean free paths below $\sim$ interatomic distances and transport heat in a random walk manner. Localized vibrations are spatially localized and are often considered to contribute negligibly to thermal transport. These categorizations have been used to understand various thermal properties in numerous amorphous solids microscopically \cite{lv_direct_2016, lv_examining_2016, lv_non-negligible_2016,larkin_thermal_2014, moon_propagating_2018, deangelis_thermal_2018, sorensen_heat_2020, aryana_tuning_2021}. Consensus is that diffusive vibrations govern thermal transport in most amorphous solids due to large degrees of disorder hindering propagation of vibrations, leading to low thermal conductivity in dielectric amorphous solids ($\lesssim 1$ Wm\textsuperscript{-1}K\textsuperscript{-1}). 

Recently, amorphous carbon has drawn immense attention due to its remarkable mechanical and thermal properties \cite{zeng_pressure_2018, lin_amorphous_2011, shang_enhancement_2023, zhang_discovery_2022, zhang_narrow-gap_2021, shang_ultrahard_2021}: Amorphous carbon with similar density to crystalline diamond (3.5 g cm\textsuperscript{-3} at 300 K) can achieve larger Young's modulus and hardness than those of crystalline diamond \cite{shang_ultrahard_2021} and it has among the highest thermal conductivity reported for pure bulk amorphous solids up to $\sim$ 37 Wm\textsuperscript{-1}K\textsuperscript{-1} at 300 K, comparable to that of many crystalline solids such as sapphire \cite{shang_enhancement_2023, cahill_thermal_1998}. Numerous experimental \cite{shamsa_thermal_2006, bullen_thermal_2000, shang_elastic_2020, shang_enhancement_2023} and computational works \cite{lv_phonon_2016, giri_atomic_2022, suarez-martinez_effect_2011, minamitani_relationship_2022} have been conducted to elucidate the nature of atomic vibrations in amorphous carbon and to explore extreme thermal properties achievable in amorphous solids.

Prior measurements have shown that thermal conductivity has a strong positive correlation with mass density and \textit{sp\textsuperscript{3}} content, demonstrating that local atomic environment could be an important factor in thermal transport in amorphous carbon \cite{shamsa_thermal_2006, bullen_thermal_2000, shang_elastic_2020, shang_enhancement_2023}. However, thermal conductivity measurements of amorphous carbon vary significantly in literature (from $\sim 1$ to $\sim 37$ Wm\textsuperscript{-1}K\textsuperscript{-1}) depending on many variables such as synthesis techniques and conditions that lead to different densities, sample qualities, and possible crystallinities. Accurate and systematic atomic simulation studies are, thus, necessary to microscopically understand the thermal transport mechanisms in amorphous carbon and these will, in larger effect, play an essential role in understanding extreme thermal properties of amorphous solids. 

Recent molecular dynamics simulations on amorphous carbon support the local structure (density and \textit{sp\textsuperscript{3}} content)-property (thermal conductivity) relationship \cite{lv_phonon_2016, giri_atomic_2022, suarez-martinez_effect_2011, minamitani_relationship_2022}. By categorizing vibrations into propagating and diffusive modes, these simulation works have further reported that diffusive vibrations play a substantial role in thermal conduction in amorphous carbon (up to 70 \% of thermal conductivity) at 300 K, a similar behavior observed in most other amorphous solids \cite{moon_sub-amorphous_2016, deangelis_thermal_2018,zhou_contribution_2017}. However, these simulations report: 1. unphysically large vibrational density of states at high frequencies above $\sim$ 70 THz similar to hydrogen vibrational frequencies using simple empirical potentials \cite{lv_phonon_2016}, 2. small simulations cells, comprised of hundreds to thousands of atoms, not able to adequately include low frequency vibrations \cite{lv_phonon_2016, giri_atomic_2022, minamitani_relationship_2022} 3. classical thermal conductivity values at room temperatures where phonon occupation is inaccurately described below Debye temperatures ($\sim$ 2300 K for amorphous carbon) \cite{giri_atomic_2022, suarez-martinez_effect_2011}. Despite these experimental and computational efforts, accurate understanding of thermal transport in amorphous carbon is still missing.

In this work, we perform molecular dynamics simulations of amorphous carbon using a machine learning potential based on neural networks. Large structures made of 110,592 atoms with varying densities from 3.0 to 3.8 g cm\textsuperscript{-3} are considered. With homogeneous non-equilibrium molecular dynamics descriptions of thermal conductivity based on a linear response theory, we decompose thermal conductivity into spectral contributions, enabling a quantum correction to the phonon occupation and heat capacity. The combination of machine learning potential with first principles accuracies, large structures, and quantum correction to heat capacity mitigates prior limitations mentioned before. By comparing with three widely used physical models and directly analyzing dispersions, we show that remarkably, propagating excitations govern thermal transport in amorphous carbon at room temperature ($\sim$ 100 \% of thermal conductivity) similar to simple crystals but unlike most amorphous solids. This efficient mode of transport may explain the high thermal conductivity values observed in amorphous carbon. Further, through temperature dependent analysis, we find that large changes in density dependent thermal conductivity observed in amorphous carbon are due to anharmonicity sensitive propagating vibrational excitations at low frequencies below $\sim$ 10 THz. Our results provide novel insights and design principles into a possible mechanism behind achieving high thermal conductivity in amorphous solids.

\section{Results and discussion}

\subsection{Amorphous carbon structures}
\begin{figure}[h!]
	\centering
	\includegraphics[width=1.0\linewidth]{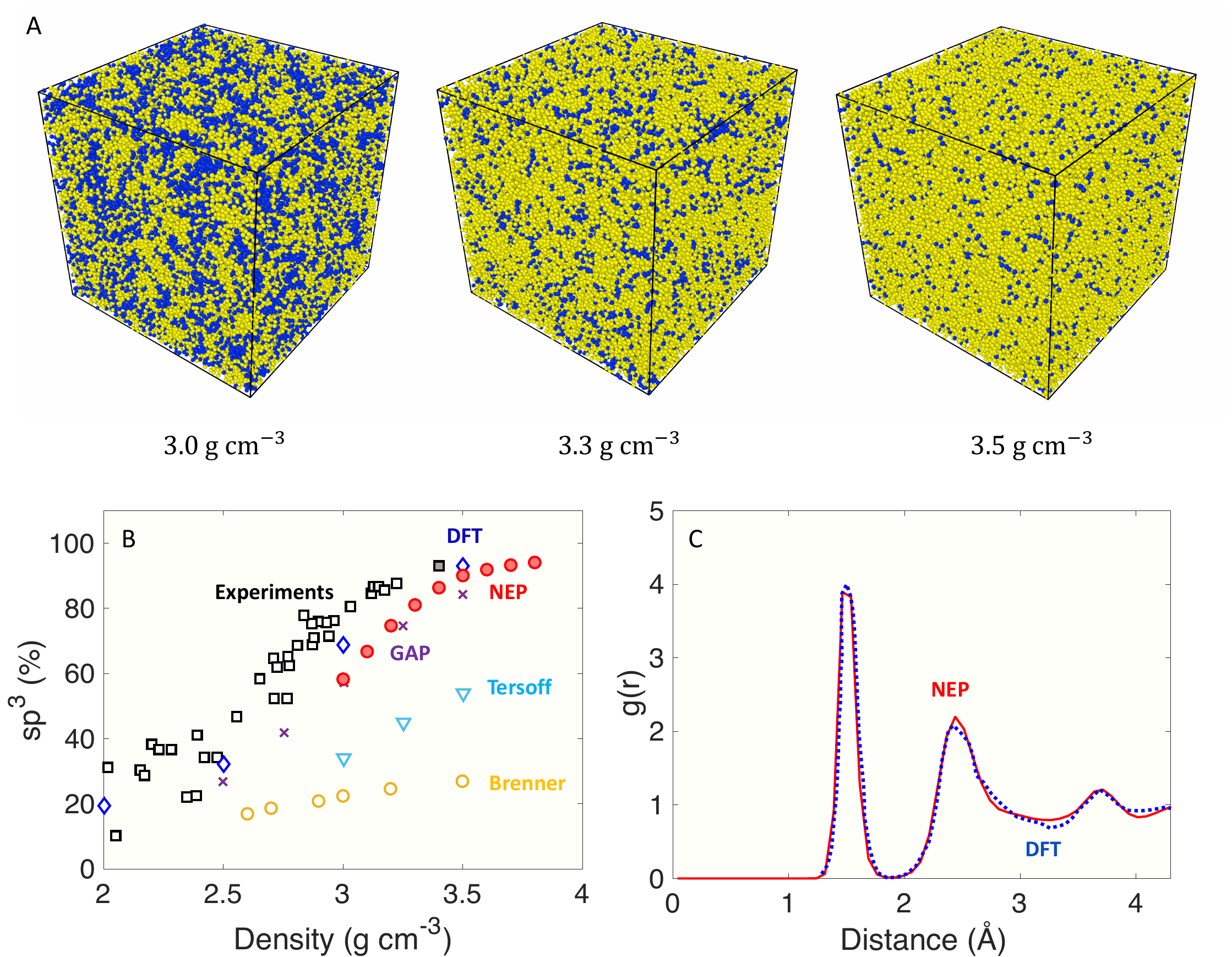}
	\caption{(A) Local atomic environment of tetrahedral amorphous carbon at different mass densities. Yellow and blue atoms denote atoms with \textit{sp\textsuperscript{3}} and \textit{sp\textsuperscript{2}} hybridizations, respectively. At 3.0, 3.3, and 3.5 g cm\textsuperscript{-3}, 58.3, 81.0, and 90.1 \% of atoms are \textit{sp\textsuperscript{3}}-bonded, respectively. (B) Density dependent \textit{sp\textsuperscript{3}} distributions of our NEP systems compared against previously reported values for various amorphous carbon systems from atomic simulations \cite{deringer_machine_2017, pastewka_describing_2008} and experiments \cite{schwan_tetrahedral_1996, fallon_properties_1993, ferrari_density_2000, shang_ultrahard_2021}.  (C) Pair distribution function of our system with 3.5 g cm\textsuperscript{-3} using the NEP potential (red curve) vs. prior literature using DFT \cite{deringer_machine_2017} (blue curve). }
	\label{fig:Fig1}
\end{figure}

Representative amorphous solid structures at 3.0, 3.3, and 3.5 g cm\textsuperscript{-3} are shown in Fig. \ref{fig:Fig1}A. Molecular dynamics simulation details are discussed in Methods. Number of atoms having four nearest neighbors (\textit{sp\textsuperscript{3}}-bonded) increase with density. A negligible number of atoms had two or five neighbors. Cutoff distance of 1.85 Å was used to determine atomic coordination numbers \cite{deringer_machine_2017, sosso_understanding_2018}. A large, spatially homogenous increase in the number of \textit{sp\textsuperscript{3}}-bonded atoms with increase in density is clearly visible. All \textit{sp\textsuperscript{3}}-bonded atoms were subsequently identified for all structures and their density dependent population is plotted against prior experimental \cite{schwan_tetrahedral_1996, fallon_properties_1993, ferrari_density_2000, shang_elastic_2020} and computational works \cite{deringer_machine_2017, pastewka_describing_2008} in Fig. \ref{fig:Fig1}B. While all data depict monotonic increase in \textit{sp\textsuperscript{3}} population with density, some empirical potentials such as Tersoff \cite{deringer_machine_2017} and Brenner \cite{pastewka_describing_2008} underpredict the \textit{sp\textsuperscript{3}} population by more than a half compared to experimental values. Structures based on machine-learning potentials such as Gaussian approximation potential (GAP) \cite{deringer_machine_2017} and NEP used here appear to have more consistent \textit{sp\textsuperscript{3}} population predictions against experiments and density functional theory (DFT) calculations. Resulting pair distribution function,  $g(r)=\frac{1}{4\pi N n r^2}\sum_{i,j} \langle \delta(r- |\boldsymbol{r}_i - \boldsymbol{r}_j|)\rangle$, is demonstrated for our 3.5 g cm\textsuperscript{-3} structure along with a prior DFT structure \cite{deringer_machine_2017} at the same density in Fig. 1C. Here, $N$ is the number of atoms, $n$ is the number density, $\boldsymbol{r}_i$ is the atomic position of the $i$th atom, and the angled brackets denotes an ensemble average. As expected for an amorphous solid, broad peaks and valleys are observed. Consistent pair distribution functions are demonstrated between the two curves.

\clearpage

\subsection{Vibrations in amorphous carbon}
\begin{figure}[h!]
	\centering
	\includegraphics[width=0.7\linewidth]{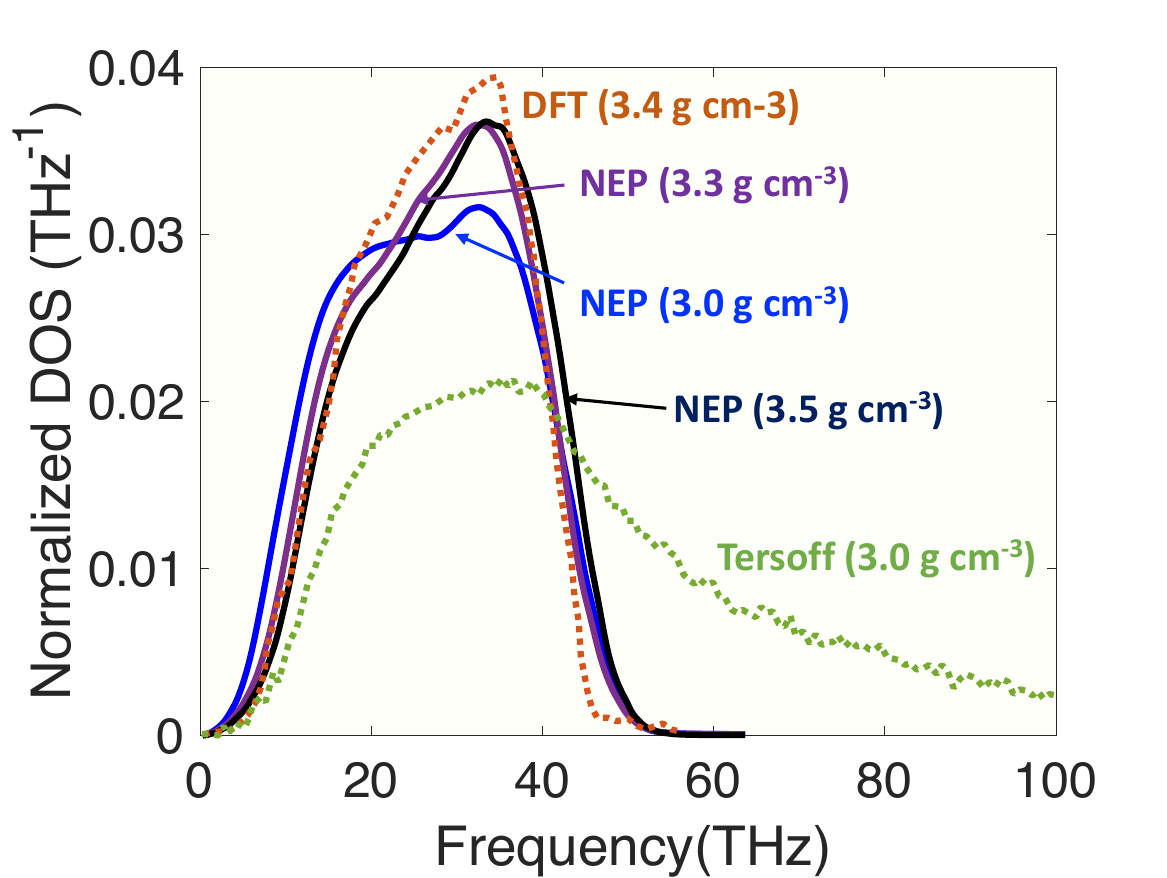}
	\caption{Vibrational densities of states (DOS) of various amorphous carbon systems. DOS for the 3.0, 3.3, and 3.5 g cm\textsuperscript{-3} structures studied here are shown as solid blue, purple, and black curves, respectively. DFT calculations (3.4 g cm\textsuperscript{-3}) are shown as a brown dotted curve and DOS using Tersoff potential (3.0 g cm\textsuperscript{-3}) is depicted as a green dotted curve.}
	\label{fig:Fig2}
\end{figure}

Using these structures, we next examine their vibrational densities of states (DOS) using spectral velocity autocorrelations (see Fig. \ref{fig:Fig2}) \cite{moon_atomic_2023, grest_density_1981, moon_heat_2024}. Due to a large sample quantity typically required for density of states measurements, experimental measurement comparison is absent here. With increase in density, we observe a flatter DOS at low frequencies below 10 THz, which signifies higher Debye sound velocities ($v_D$) as DOS scales as $v_D^{-3}$. By fitting the Debye model of vibrational density of states below 5 THz, we obtain 12.97, 14.60, and 15.21 km s\textsuperscript{-1} for the 
3.0, 3.3, and 3.5 g cm\textsuperscript{-3} structures, respectively. Vibrational densities of states using the NEP potential appear to be more consistent with prior DFT calculations \cite{sosso_understanding_2018} while the DOS using a Tersoff potential \cite{lv_phonon_2016} predicts a significant number of vibrations with frequencies higher than 50 THz. With the NEP potential, we then have structural and vibrational properties with first-principles accuracies for our large amorphous carbon structures.

\subsection{Thermal conductivity and transport mechanisms}

Thermal conductivity values for all the structures were then determined by homogeneous non-equilibrium molecular dynamics (HNEMD) \cite{fan_homogeneous_2019, gabourie_spectral_2021} (See Methods). Similar to the Green-Kubo (GK) formalism \cite{kubo_fluctuation-dissipation_1966} in equilibrium molecular dynamics (EMD), the HNEMD method is based on a linear response theory. In HNEMD, independent phonon calculations are not required to obtain spectral thermal conductivity unlike other molecular dynamics based methodologies such as Green-Kubo modal analysis (GKMA) \cite{lv_direct_2016} and normal mode decomposition (NMD) \cite{mcgaughey_predicting_2014} which limit the number of atoms due to large computational costs of constructing and diagonalizing dynamical matrices. 

\begin{figure}[h!]
	\centering
	\includegraphics[width=1\linewidth]{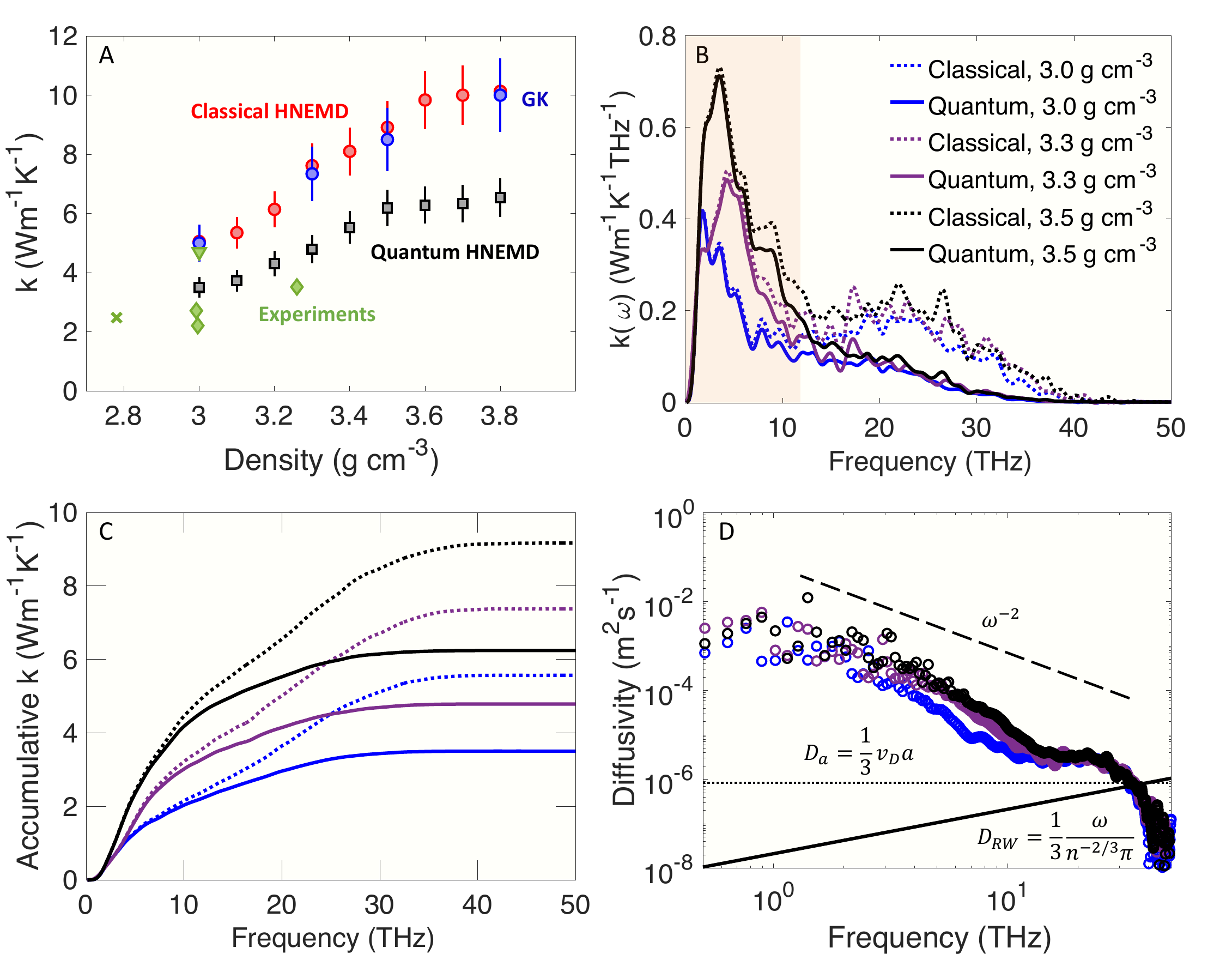}
	\caption{(A) Density dependent classical (red solid circles) and quantum mechanical (black solid squares) thermal conductivity at 300 K compared against thermal conductivity values using Green-Kubo formalism (blue solid circles) and available measurements (green symbols). Sample synthesis technique and thickness are: cross (filtered arc, 47 nm) \cite{bullen_thermal_2000}, diamond (filtered cathodic vacuum arc, 18.5 to 100 nm) \cite{shamsa_thermal_2006}, and inverse triangle (filtered cathodic vacuum arc, 20 to 100 nm) \cite{chen_thermal_2000} (B) Classical (solid curves) and quantum mechanical (dotted curves) spectral thermal conductivity. Shaded orange region highlights large density dependence in spectral thermal conductivity. (C) Thermal conductivity accumulation function of 3.0, 3.3, and 3.5 g cm\textsuperscript{-3} structures at 300 K. (D) Spectral diffusivity of 3.0, 3.3, and 3.5 g cm\textsuperscript{-3} structures at 300 K. Color schemes in (C) and (D) are the same as (B). Dashed line representing inverse quadratic power law is a guide to the eye. Dotted line represents diffusivity $D_a$ when the mean free path is interatomic distance. Solid black line depicts a diffusivity based on a random-walk theory $D_{RW}$.  }
	\label{fig:Fig3}
\end{figure}

Classical and quantum corrected thermal conductivity of all systems studied here were plotted against experimental measurements and Green-Kubo thermal conductivity independently calculated here as shown in Fig. \ref{fig:Fig3}. Green-Kubo thermal conductivities were obtained by $k = \frac{V}{3k_BT}\int_0^\infty dt \langle \boldsymbol{J}(t) \cdot \boldsymbol{J}(0) \rangle$ where $\boldsymbol{J}(t)$ is the heat current. Classical thermal conductivity from HNEMD and GK thermal conductivity are within the errorbars of each other. There have been a lot of variations in the measured thermal conductivities for amorphous carbon depending on the synthesis techniques, thickness, and sample quality. Nonetheless, we observe that quantum corrected thermal conductivity for the amorphous carbon structures discussed here is on the same order of magnitude as measurements at 300 K \cite{bullen_thermal_2000, shamsa_thermal_2006, chen_thermal_2000}.

To see the effect of the quantum correction to the heat capacity on the total thermal conductivity in amorphous carbon, spectral thermal conductivity and thermal conductivity accumulation functions of 3.0, 3.3, and 3.5 g cm\textsuperscript{-3} structures at 300 K are shown in Fig. \ref{fig:Fig3}B and \ref{fig:Fig3}C, respectively. It appears that the effect of quantum correction to the specific heat on thermal conductivity becomes prominent above $\sim$ 7 THz. This is expected as $k_BT$ $\sim$ 25 meV or 6 THz at room temperature. At 300 K, the quantum correction to the heat capacity reduces the classical thermal conductivities by as much as 30 to 40 \%, demonstrating the importance of considering quantum effects in these materials. Further, it is interesting to note that density affects spectral thermal conductivity of amorphous carbon drastically below $\sim$ 10 THz while there is little to no effect above 10 THz. Therefore, our work reveals that density dependent thermal conductivity in amorphous carbon widely observed in literature at room temperature is likely due to changes in transport properties of these low frequency vibrations. 

\subsubsection{Characterization of popagating and diffusive vibrations}
We next examine spectral thermal diffusivity values for the 3.0, 3.3, and 3.5 g cm\textsuperscript{-3} structures at 300 K obtained by $D(\omega)=\frac{k(\omega)}{C(\omega)DOS(\omega)}$ as plotted in Fig. \ref{fig:Fig3}D along with some widely used diffusivity models for amorphous materials. There currently exist a few different methods to distinguish propagating vibrations and diffusive vibrations: some methods require solving the dynamical matrices and subsequently examine normal mode shapes or properties (e.g. eigenvector periodicity \cite{seyf_method_2016}) and the other methods rely on physical models that describe propagating and diffusive vibrations. The first route is not computationally feasible here. Our systems ($\sim$ 10\textsuperscript{5} atoms) are more than an order of magnitude larger than typical numbers of atoms used for dynamical matrix calculations using simple empirical potentials in literature (10\textsuperscript{3} to 10\textsuperscript{4} atoms) \cite{larkin_thermal_2014, giri_atomic_2022, moon_propagating_2018}. Therefore, we resort to physical models describing propagating and diffusive vibrations. 

The first model relies on how propagating wave lifetimes depend on frequency. At THz frequencies, lifetimes of propagating vibrations typically follow $\omega^{-2}$. Prior works determined the frequency at which $\omega^{-2}$ dependence in lifetimes and diffusivity disappears as the transition frequency from propagating to diffusive vibrations in amorphous solids including amorphous carbon \cite{larkin_thermal_2014, zhou_contribution_2017, lv_phonon_2016}. Using this criterion, we find that the transition frequency can be approximated to be around $\sim$ 32 THz for these amorphous carbon structures (See Fig. S1 for more details in Supplementary Information). 

The second physical model considers the lowest diffusivity ($D_a$) possible by a propagating wave where group velocity is sound velocity (here, we use the Debye velocity of $\sim$ 14 km s\textsuperscript{-1} as an average value) and mean free path is the interatomic distance, $a$ \cite{kittel_interpretation_1949}. The third model considered here considers maximum diffusivity for diffusive vibrations and is derived from a random walk theory \cite{agne_minimum_2018} that has a linear dependence on frequency as $D_{RW}=\frac{1}{3}  \frac{\omega}{\pi n^{2/3}}$ where n is the number density. If physically realistic, these two independent models should ideally merge at a certain frequency for amorphous carbon. These two criteria have been previously used in amorphous solids but also in complex crystals to determine propagating vs. diffusive normal modes that resulted in consistent thermal conductivity with experiments \cite{luo_vibrational_2020, moon_propagating_2018, kim_origin_2021, cai_diffuson-dominated_2023}. Applying these two physical models lead to nearly identical crossover frequencies at remarkably high $\sim$ 35 THz in amorphous carbon as shown in Fig. \ref{fig:Fig3}D. 

\begin{figure}[h!]
	\centering
	\includegraphics[width=0.47\linewidth]{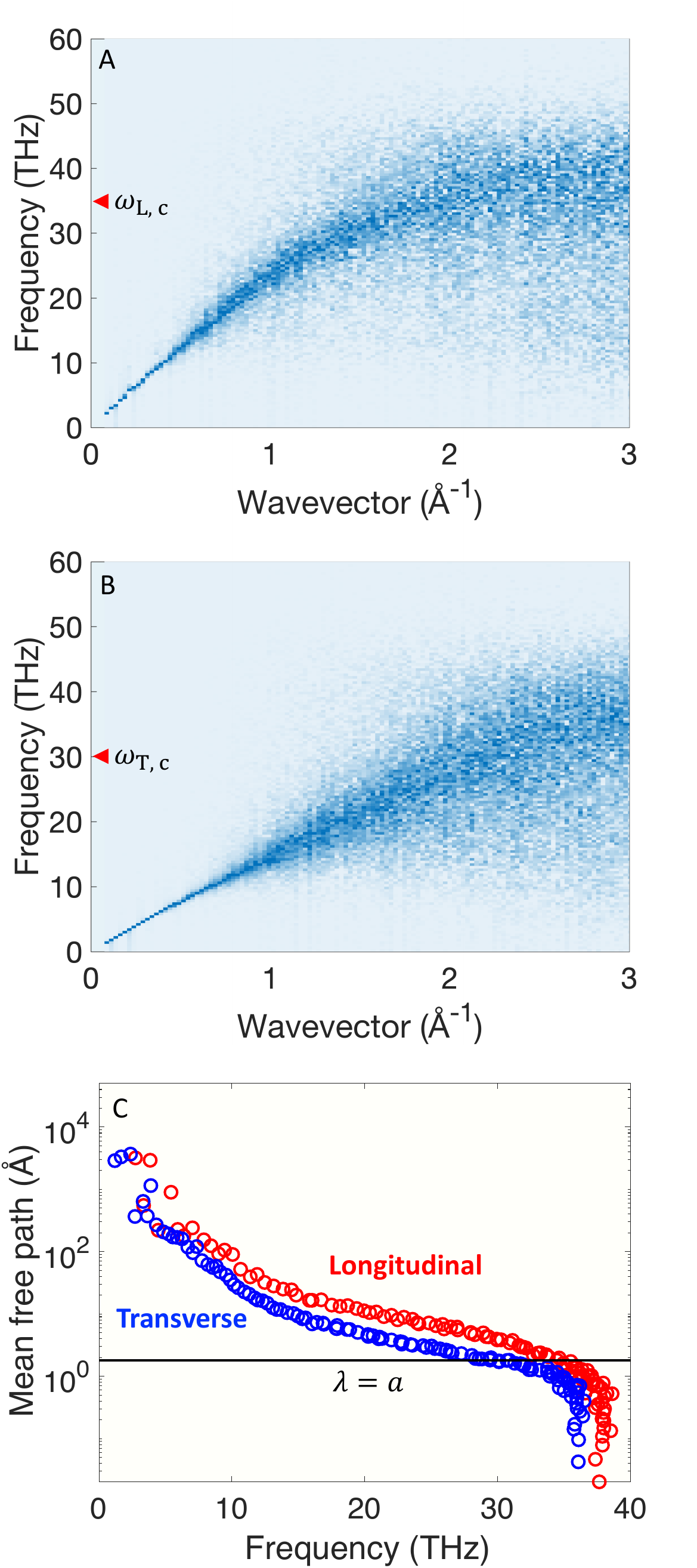}
	\caption{Wavevector and frequency resolved (A) longitudinal and (B) transverse velocity current correlations for the 3.5 g cm\textsuperscript{-3} structure. Estimated crossover frequencies from propagating to diffusive vibrations are noted as $\omega_{L, c}$ and $\omega_{T, c}$ for longitudinal and transverse directions, respectively. (C) Extracted mean free paths ($\lambda$) for both longitudinal (red circles) and transverse (blue circles) excitations. Horizontal black line is a guide to eye representing $\lambda$ approaching interatomic distance ($a$) of amorphous carbon.}
	\label{fig:Fig5}
\end{figure}

Three physical models considered here consistently lead to remarkably high crossover frequencies of $\sim$ 30 to 35 THz. For these propagating vibrational excitations then, crisp dispersions with well-defined frequencies, group velocities, and mean free paths should be observed \cite{monaco_anomalous_2009, shintani_universal_2008, moon_thermal_2019, sette_dynamics_1998, fiorentino_hydrodynamic_2023}. Representative longitudinal and transverse dispersions for the 3.5 g cm\textsuperscript{-3} amorphous carbon structure are shown in Fig. \ref{fig:Fig5}A and B (see Methods for calculation details). At first glance, clear dispersions are demonstrated up to $\sim$ 30 THz and $\sim$ 20 THz for longitudinal and transverse excitations, respectively. By carefully fitting the single damped harmonic oscillator model at each wavevector shown in Fig. S2 in the Supplementary Information, we report mean free paths of longitudinal and transverse excitations in Fig. \ref{fig:Fig5}C. We find that propagating to diffusive crossover frequencies for longitudinal and transverse excitations are in a similar range as the other physical models at $\sim$ 30 and 35 THz, respectively. Further, the mean free paths vary by nearly four orders of magnitude from $\sim$ 1 THz to the crossover frequencies, similar to spectral thermal diffusivity calculated by HNEMD as shown in Fig. \ref{fig:Fig3}D. 

All evidences based on the three physical models and dispersion analysis are pointing to the consequence that despite having intrinsically disordered structures, propagating vibrations that exist below 30 to 35 THz are responsible for thermal transport ($\sim$ 100 \% of thermal conductivity) in amorphous carbon at 300 K as demonstraed in Fig. \ref{fig:Fig3}C unlike most amorphous solids reported in literature where diffusive vibrations dominate thermal conduction \cite{deangelis_thermal_2018, wingert_thermal_2016}. This behavior is remarkably similar to thermal transport in simple dielectric crystals. Our finding is in contrast with prior molecular dynamics predictions of amorphous carbon where a substantial contribution from diffusive vibrations is reported (up to 70 \% of thermal conductivity) \cite{giri_atomic_2022, lv_phonon_2016}. We attribute this difference to a combination of three possible factors: 1. Simple empirical potentials may lack accuracy in predicting vibrational properties \cite{giri_atomic_2022, lv_phonon_2016, suarez-martinez_effect_2011}, 2. Prior simulations typically comprise of hundreds or thousands of atoms, not being able to include vibrations with long wavelengths and small frequencies adequately \cite{giri_atomic_2022, lv_phonon_2016, minamitani_relationship_2022}, and 3. Due to the lack of spectral decomposition of thermal conductivity, a quantum correction to the heat capacity was not possible, leading to overpredictions of high frequency mode contributions \cite{giri_atomic_2022}.


\clearpage

\subsubsection{Temperature dependent thermal transport}
\begin{figure}[h!]
	\centering
	\includegraphics[width=1\linewidth]{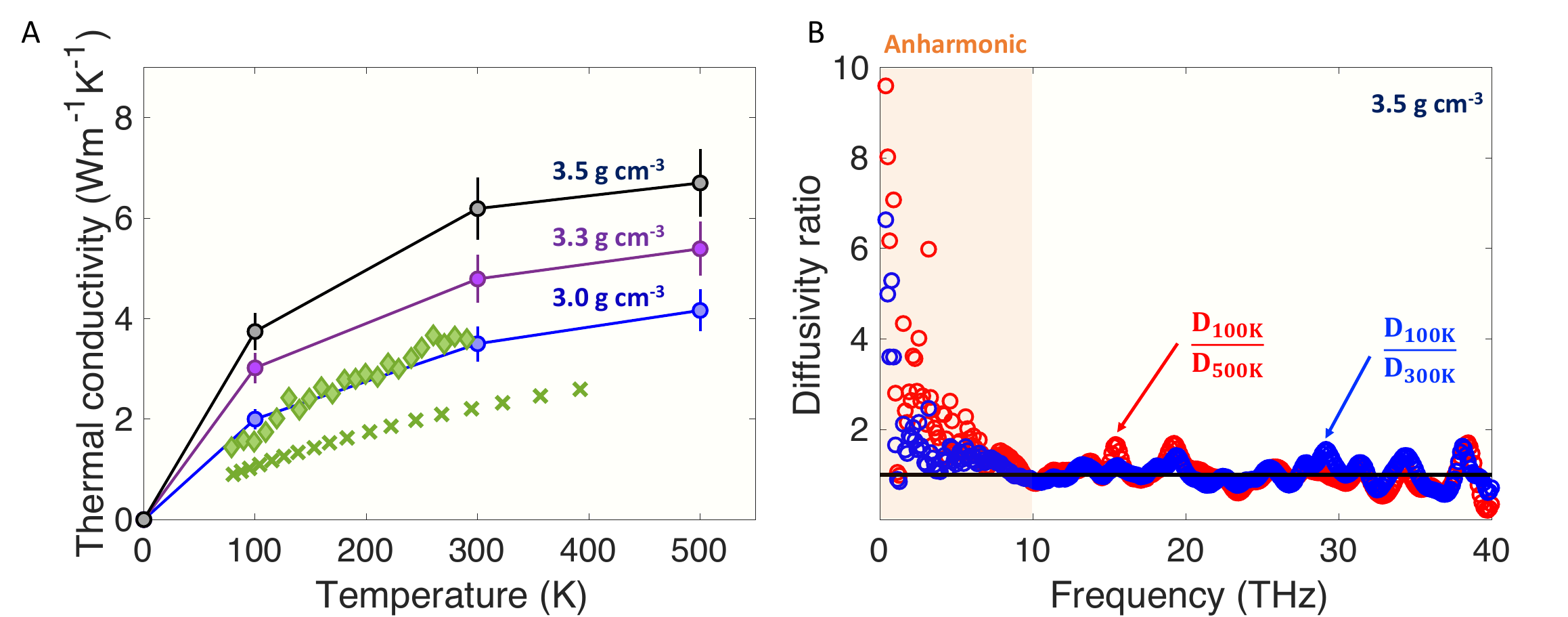}
	\caption{(A) Temperature dependent thermal conductivity for 3.0 (blue solid circles), 3.3 (purple solid circles), and 3.5 g cm\textsuperscript{-3} (black solid circles) structures compared against prior measurements (green solid diamond: filtered cathodic vacuum arc, 3.3  g cm\textsuperscript{-3} \cite{shamsa_thermal_2006} and green cross: filtered arc, 2.8  g cm\textsuperscript{-3} \cite{bullen_thermal_2000}). These data and density dependent thermal conductivity data with the same symbols are from the same prior studies. (B) Spectral diffusivity ratios, $D_{100 K}/D_{500 K}$  (red open circles) and $D_{100 K}/D_{300 K}$  (blue open circles). Diffusivity ratio of unity is marked as black holizontal line as a guide to the eye. }
	\label{fig:Fig4}
\end{figure}

So far, we have considered how thermal occupation affects thermal conductivity and the role of propagating vibrations in amorphous carbon at 300 K. We next explore the physical nature of scattering mechanisms of these propagating vibrations through examining temperature dependent thermal transport at 100, 300, and 500 K as shown in Fig. \ref{fig:Fig4}. Only the quantum corrected thermal conductivity values are plotted in Fig. \ref{fig:Fig4}A along with some prior measurements \cite{bullen_thermal_2000, shamsa_thermal_2006}. Similar to density dependent thermal conductivity (see Fig. 3A), some differences between our predictions and measurements may arise from a variety of both experimental and computational origins including deposition techniques, sample thicknesses, and accuracy of the interatomic potential used here. A similar order of magnitude in thermal conductivity among the measurements and predictions is, nonetheless, shown. In all systems, a typical glass-like thermal conductivity temperature dependence is observed: thermal conductivity increases with increase in temperature. 

Spectral diffusivity varies by $~\sim$ 6 orders of magnitude (see Fig. 3D). Therefore, diffusivity ratios for 100 K, 300 K, and 500 K are used as our metric of temperature effect for clear visualization. These ratios are representatively plotted for the 3.5 g cm\textsuperscript{-3} structure in Fig. \ref{fig:Fig4}B. Results for other amorphous carbon systems considered in this work are generally consistent with the 3.5 g cm\textsuperscript{-3} system and are not shown for redundancy. Above $\sim$ 10 THz, both ratios fluctuate around unity, demonstrating that scattering mechanisms above $\sim$ 10 THz are nearly temperature invariant and may originate from structural disorder in amorphous carbon. Below $\sim$ 10 THz, we observe more significant temperature effects in the spectral diffusivity. Generally, $D_{100 K}/D_{500 K}$  is greater than $D_{100 K}/D_{300 K}$ in this frequency range, indicating that diffusivity monotonically decreases with increase in temperature. 
We attribute the temperature dependence of thermal diffusivity to anharmonic quasi-particle interactions in this frequency range. Similar trends are observed in independently calculated temperature dependent dispersion linewidths shown in Fig. S3. The trend that the ratios increase as frequency decreases is thought to originate from vibrations with larger periods (hence, larger wavelengths) becoming less sensitive to local disorders.

Our findings of the transition of scattering mechanisms from anharmonicity sensitive regime to disorder dominated regime are consistent with prior experimental works on various amorphous solids ranging from network systems \cite{kim_origin_2021, baldi_anharmonic_2014, masciovecchio_evidence_2006, ruocco_high-frequency_2001, ruffle_glass-specific_2006} to polymers \cite{sette_collective_1995, masciovecchio_inelastic_2004}: At low frequencies below $\sim$ 10 to 100 GHz, picosecond acoustics and Brillouin light scattering have shown strong temperature dependent mean free paths of acoustic excitations while negligible temperature dependence in the inelastic peak widths above peak frequency of ~ 1 THz from inelastic X-ray and neutron scattering measurements have been reported. The high transition frequency of scattering mechanisms at $\sim$ 10 THz observed here in amorphous carbon may be due to relatively high sound velocities \cite{moon_thermal_2019} from strong C-C bonds. At a given wavevector, higher frequency acoustic waves are supported in amorphous carbon vs. those with low sound velocities.

Based on our density and temperature dependent vibrational and thermal characterizations, a novel insight into how density affects microscopic thermal transport in amorphous carbon emerges. It was observed in Fig. \ref{fig:Fig2} that the enhancement in thermal conductivity with increase in density in amorphous carbon is mostly due to low frequency propagating vibrations below $\sim$ 10 THz. From our temperature dependent diffusivity analysis, we further conclude that more specifically, anharmonicity sensitive propagating vibrations are responsible for variations in thermal conductivity in amorphous carbon due to different densities studied here. We expect that our findings here can be extended to other amorphous solids, especially those with local tetrahedral orders including amorphous silicon, germanium, and silica. Direct verifications of our predictions may be possible through inelastic X-ray or neutron scattering measurements of vibrational density of states and dispersion of amorphous carbon. 

\section{Conclusions}

Using a machine learning potential based on neural networks for large-scale amorphous carbon structures in molecular dynamics, we elucidate the microscopic mechanism behind high thermal conductivity values in amorphous carbon, among the highest in amorphous solids. Through spectral decomposition of thermal conductivity, quantum correction to the heat capacity, and dispersion analysis, we demonstrate that propagating vibrational excitations, present up to 30 $\sim$ 35 THz, dominate thermal transport in amorphous carbon at room temperature ($\sim$ 100 \% of thermal conductivity). Further, temperature dependent analysis of thermal diffusivity and dispersion linewidths indicate that the origin of density dependent thermal conductivity is due to anharmonicity sensitive propagating vibrational excitations at low frequencies below $\sim$ 10 THz, making connections between structure and property. Engineering anharmonicity sensitive vibrations could be key to achieving amorphous solids with high thermal conductivity. We have suggested an experimental approach based on inelastic scattering to test these predictions. 

\section{Methods}

\subsection{Molecular dynamics simulations}
Crystal diamond of 110,592 atoms at various densities from 3.0 to 3.8 g cm\textsuperscript{-3} were melted at 15,000 K for 30 ps, followed by quenching to an amorphous state at 1000 K at 100 K ps\textsuperscript{-1} in NVT ensemble (constant number of atoms, volume, and temperature) using Graphics Processing Units Molecular Dynamics (GPUMD) \cite{fan_GPUMD_2022}. Resulting structures were then annealed for 1 ns at 1000 K and were subsequently cooled to 300 K at 100 K ps\textsuperscript{-1} in NVT. Using these seed structures, molecular dynamics outputs were recorded for 2 ns after initial equilibration of 1 ns at desired temperatures (100, 300, and 500 K) under NVT. Our large structures result in more statistics of propagating vibrational excitations compared to prior molecular dynamics simulations with thousands of atoms \cite{lv_phonon_2016, giri_atomic_2022}. A timestep of 1 fs was used. Three runs with different initial velocities were carried out for better statistics for thermal and vibrational property calculations. Neuro-evolution potential (NEP), a neural network based machine-learning potential, was utilized to describe interatomic interactions with first-principles accuracies \cite{fan_GPUMD_2022, fan_neuroevolution_2021}. 

\subsection{Thermal conductivity calculations}

In HNEMD, one applies an external driving force perturbation and examines the materials response (heat current) to it as
\begin{equation}
    \boldsymbol{F}_i^{ext} = \boldsymbol{F}_e \cdot \boldsymbol{W}_i
    \label{eq: ext_F}
\end{equation}
where $\boldsymbol{F}_e$ is the driving force parameter with the dimension of inverse length, $\boldsymbol{W}_i = \sum_{j\neq i}\frac{\partial U_j}{\partial \boldsymbol{r}_{ij}}\otimes \boldsymbol{r}_{ij}$ is the 3 by 3 virial tensor of an atom, $\otimes$ represents a tensor product and $U_j$ is the atomic potential energy. A small value of $\boldsymbol{F}_e$ at $2 \times 10^{-4}$ \AA\textsuperscript{-1} was used to ensure the system is in the linear response regime. 

Instantaneous heat current is related to the virial tensor by
\begin{equation}
    \boldsymbol{J}(t) = \sum_i \boldsymbol{W}_i \cdot \boldsymbol{v}_i.
    \label{eq: inst_heat}
\end{equation}
Ensemble average of the heat current is then directly related to the driving force parameter and thermal conductivity by
\begin{equation}
    \langle J^\alpha \rangle = TV \sum_\beta k^{\alpha\beta}F_e^\beta
\label{eq:avg_heat}
\end{equation}
where $\alpha$ and $\beta$ represent Cartesian directions, $T$ is temperature, and $V$ is system volume. Thermal conductivity can therefore be determined from Eq. \ref{eq: ext_F}-\ref{eq:avg_heat}. For isotropic systems like amorphous carbon, isotropic thermal conductivity is expected. Therefore, we generally refer to thermal conductivity as $k$. We can further decompose thermal conductivity into spectral contributions through virial-velocity function $Q(t) = \sum_i \langle \boldsymbol{W}_i(0) \cdot \boldsymbol{v}_i(t) \rangle$ as 
\begin{equation}
    k(\omega) = \frac{2}{VTF_e}\int_{-\infty}^{\infty}Q(t)e^{i\omega t}dt.
\end{equation}
Due to the classical nature of molecular dynamics simulations, atomic dynamics follows the Maxwell-Boltzmann distribution. With spectral thermal conductivity, one could, therefore, apply a spectral `quantum correction' to the occupation number and heat capacity such that 
\begin{equation}
    k_Q(\omega) = k(\omega) \frac{x^2e^x}{(e^x-1)^2}.
\end{equation}
where $x = \frac{\hbar\omega}{k_BT}$, $\hbar$ is the reduced Planck constant, and $k_B$ is the Boltzmann constant \cite{wang_quantum-corrected_2023}. HNEMD and spectral decompositions of thermal conductivity have been utilized in various materials including graphene \cite{gabourie_spectral_2021} and amorphous silicon \cite{wang_quantum-corrected_2023}.

\subsection{Dispersion analysis}
Longitudinal and transverse dispersions from wavevector and frequency resolved velocity current correlations for amorphous carbon were obtained by \cite{monaco_anomalous_2009, shintani_universal_2008, moon_propagating_2018, moon_thermal_2019}
\begin{equation}
    C_{L,T}(q,\omega) = \frac{q^2}{2\pi\omega^2N} \int dt \big \langle \boldsymbol{j}_{L,T}(q,t) \cdot \boldsymbol{j}_{L,T}(-q,0) \big \rangle e^{i\omega t}
\end{equation}
where subscripts, $L, T$, refer to longitudinal or transverse branches, $q$ is the wavevector, $\omega$ is the radial frequency, and $\boldsymbol{j}_{L,T}(q,t)$ is given by
\begin{equation}
    \boldsymbol{j}_L(q,t) = \sum_i^N \big (\boldsymbol{v}_i(t)\cdot \hat{\boldsymbol{q}} \big ) \hat{\boldsymbol{q}} e^{i\boldsymbol{q}\cdot \boldsymbol{r}_i(t)}
\end{equation}
\begin{equation}
    \boldsymbol{j}_T(q,t) = \sum_i^N \big (\boldsymbol{v}_i(t)-\big (\boldsymbol{v}_i(t)\cdot \hat{\boldsymbol{q}} \big ) \hat{\boldsymbol{q}} \big ) e^{i\boldsymbol{q}\cdot \boldsymbol{r}_i(t)}
\end{equation}
with $\boldsymbol{v}_i(t)$, $\boldsymbol{r}_i(t)$, and $\hat{q}$ representing atomic velocities, positions, and unit wavevector, respectively.  

\section{Acknowledgement}
Z.T. acknowledges the support of the Department of the Navy, Office of Naval Research under ONR award number N00014-22-1-2357. We are grateful for discussions with Dr. Lucas Lindsay. This work used the Extreme Science and Engineering Discovery Environment (XSEDE) Expanse under Allocation No. TG-MAT200012. This research used resources of the National Energy Research Scientific Computing Center (NERSC), a U.S. Department of Energy Office of Science User Facility located at Lawrence Berkeley National Laboratory, operated under Contract No. DE-AC02-05CH11231 using NERSC award BES-ERCAP0023621.

\section{Author contributions}
J.M. conceived the research. J.M. performed simulations and data analysis. J.M. and Z.T. interpreted the results. J.M. wrote the paper with contributions from Z.T.

\section{Competing interests}
The authors declare no competing interests.

\clearpage


\clearpage

\title{Supplementary Information for\\
Crystal-like thermal transport in amorphous carbon}

\author{Jaeyun Moon}
\email{To whom correspondence should be addressed; E-mail: jaeyun.moon@cornell.edu}
 \affiliation{Sibley School of Mechanical and Aerospace Engineering, Cornell University}

\author{Zhiting Tian}
 \affiliation{Sibley School of Mechanical and Aerospace Engineering, Cornell University}

\date{\today}

\maketitle
\clearpage

\renewcommand{\figurename}{Fig.}
\setcounter{section}{0}
\renewcommand{\thesection}{\Roman{section}} 
\renewcommand{\thefigure}{S\arabic{figure}}
\setcounter{figure}{0}

\section{Determination of transition frequency from propagating to diffusing vibrations using the frequency power law}

\begin{figure}[h!]
	\centering
	\includegraphics[width=1\linewidth]{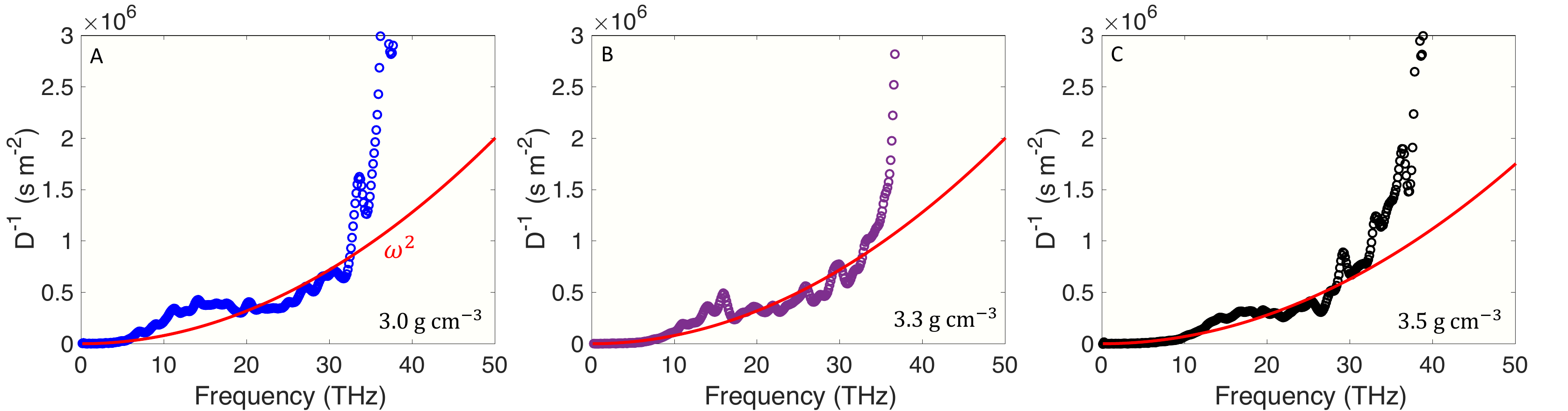}
	\caption{Inverse of spectral diffusivity with frequency for the 3.0, 3.3, and 3.5 g cm\textsuperscript{-1} amorphous carbon structures studied.}
	\label{fig:S1}
\end{figure}

As discussed by prior works, determining the transition frequency through $D \sim \omega^{-2}$ is not exact and can only give us a ball park estimation. Some fluctuations in the diffusivities are demonstrated in our systems but it is clear that above $\sim$ 32 THz, diffusivity no longer follows the power law and diverges as shown in Fig. \ref{fig:S1}. 

\section{Comment on Ref. \cite{shamsa_thermal_2006} and \cite{lv_phonon_2016}}

Temperature dependent thermal conductivity measurements from Ref. \cite{shamsa_thermal_2006} shown as green solid diamond symbols in Fig. 4 of the main text have been cited to have the mass density of 3.0 g cm\textsuperscript{-3} in Ref. \cite{lv_phonon_2016}. With this density, our thermal conductivity predictions at 3.0 g cm\textsuperscript{-3} (blue solid circles) in Fig. 4A are consistent with these measurements. However, it is our interpretation of the measurements from Ref. \cite{shamsa_thermal_2006} that the reported mass density is $\sim$ 3.3 g cm\textsuperscript{-3}. 

\section{Velocity current correlations analysis}

\begin{figure}[h!]
	\centering
	\includegraphics[width=1\linewidth]{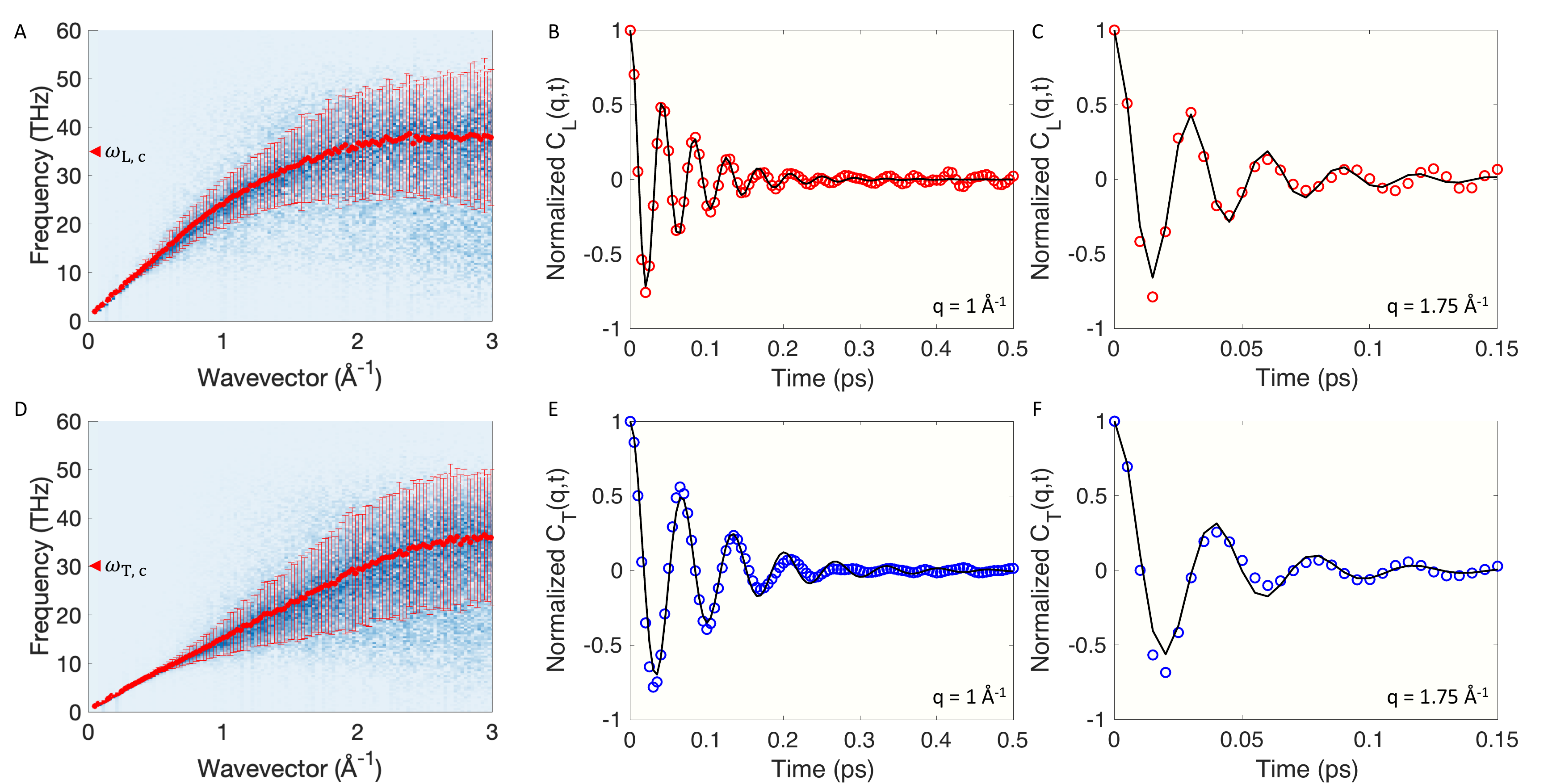}
	\caption{Wavevector and frequency resolved (A) longitudinal and (D) transverse velocity current correlations for the 3.5 g cm\textsuperscript{-3} structure. Red solid circles with the errorbars represent peak frequencies and linewidths by fitting single damped harmonic oscillator. Clear dispersions are demonstrated up to $\sim$ 30 THz in both longitudinal and transverse directions. Time resolved longitudinal velocity current correlations are shown at (B) 1 and (C) 1.75 nm\textsuperscript{-1}. Time resolved transverse velocity current correlations are shown at (E) 1 and (F) 1.75 nm\textsuperscript{-1}. Black curves are single damped harmonic oscillator model fits. Well-defined exponentially decaying oscillations are demonstrated at these wavevectors for both longitudinal and transverse directions.}
	\label{fig:S2}
\end{figure}

Time Fourier transforms ($C_L(q,t)$ and $C_T(q,t)$) of the wavevector and frequency resolved velocity current correlation functions were fit by a single damped harmonic oscillator model, $C_{L,T}(q,t) = C_{L,T}(q,0)e^{-\Gamma t/2}\cos{\omega t}$. Well-defined frequencies and relaxation times are demonstrated up to very high frequencies of $\sim$ 30 to 35 THz as shown in Fig. \ref{fig:S2}. 

\begin{figure}[h!]
	\centering
	\includegraphics[width=0.55 \linewidth]{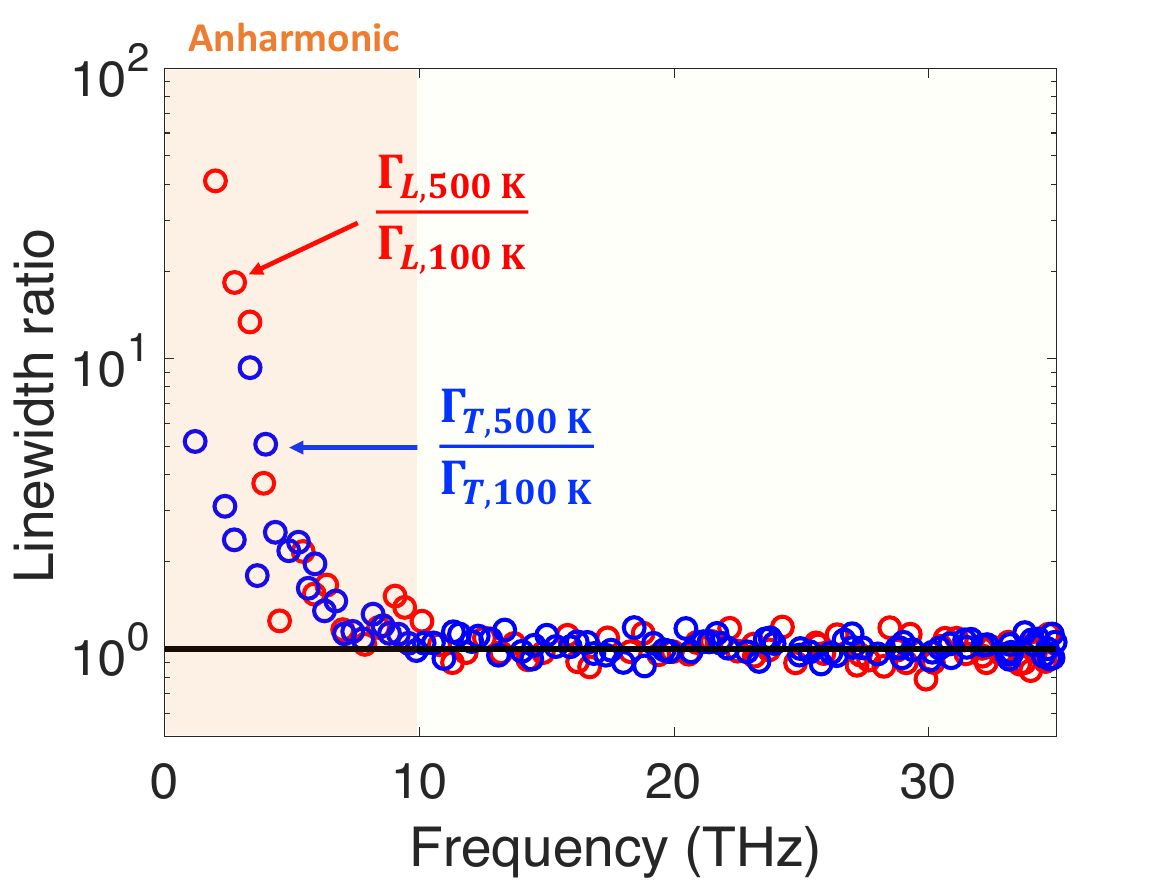}
	\caption{Linewidth ratios for both longitudinal and transverse excitations demonstrate strong temperature dependence below $\sim$ 10 THz for the 3.5 g cm\textsuperscript{-1} amorphous carbon structure. }
	\label{fig:S3}
\end{figure}

Temperature dependent analysis of dispersion linewidths depict anharmonicity sensitive propagating excitations below $\sim$ 10 THz as shown in Fig. \ref{fig:S3}, consistent with our thermal diffusivity analysis in the main text.

\clearpage



\bibliography{apssamp.bib}

\end{document}